*Machine learning*

# AMP$_0$: Species-Specific Prediction of Antimicrobial Peptides using Zero and Few Shot Learning

Sadaf Gull[1] and Fayyaz Minhas[2,*]

[1,2] PIEAS Biomedical Informatics Lab, Pakistan Institute of Engineering and Applied Sciences, PO Nilore, Islamabad, Pakistan.

[2]Department of Computer Science, University of Warwick, Coventry, UK

[1]sadafzakarkhan@gmail.com, [2]fayyaz.minhas14@alumni.colostate.edu

*To whom correspondence should be addressed.

**Abstract**

**Motivation:** The evolution of drug-resistant microbial species is one of the major challenges to global health. The development of new antimicrobial treatments such as antimicrobial peptides needs to be accelerated to combat this threat. However, the discovery of novel antimicrobial peptides is hampered by low-throughput biochemical assays. Computational techniques can be used for rapid screening of promising antimicrobial peptide candidates prior to testing in the wet lab. The vast majority of existing antimicrobial peptide predictors are *non-targeted* in nature, i.e., they can predict whether a given peptide sequence is antimicrobial, but they are unable to predict whether the sequence can target a particular microbial species.

**Results:** In this work, we have developed a targeted antimicrobial peptide activity predictor that can predict whether a peptide is effective against a given microbial species or not. This has been made possible through zero-shot and few-shot machine learning. The proposed predictor called AMP$_0$ takes in the peptide amino acid sequence and any N/C-termini modifications together with the genomic sequence of a target microbial species to generate targeted predictions. It is important to note that the proposed method can generate predictions for species that are not part of its training set. The accuracy of predictions for novel test species can be further improved by providing a few example peptides for that species. Our computational cross-validation results show that the proposed scheme is particularly effective for targeted antimicrobial prediction in comparison to existing approaches and can be used for screening potential antimicrobial peptides in a targeted manner especially for cases in which the number of training examples is small.

**Availability:** We have also developed a webserver of the proposed methodology available at http://ampzero.pythonanywhere.com. The data used for training and testing is also available for downloading, given in supplementary material.

**Contact:** fayyaz.minhas14@alumni.colostate.edu

**Supplementary information:** Supplementary data are available.

## 1 Introduction

Antibiotics play a significant role in protecting humans from microbial infections. The discovery and use of antibiotics since the 1930s has helped in treating serious infections and saved many lives (Aslam *et al.*, 2018). (Blair, 2018; Ventola, 2015). Resistance against antibiotics in microbes was detected in the 1960s and it prompted an evolutionary arms race between microbes and antibiotics (Ventola, 2015). Antimicrobial resistance is currently a major global health crisis. The number of deaths due to infections caused by antibiotic resistance annually is increasing and is estimated to reach up to 10 million by 2050 (Blair, 2018).

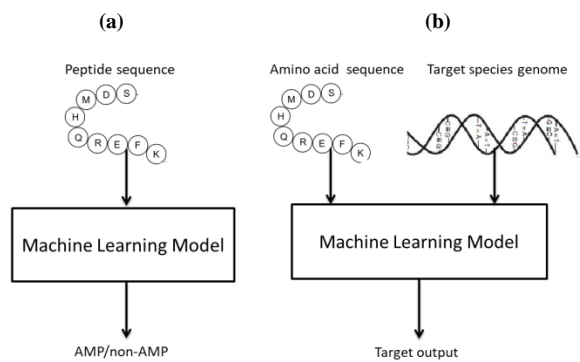

**Fig. 1.** A general framework of machine learning predictors for (a) non targeted and (b) targeted predictions

World Health Organization (WHO) has generated a list of antibiotic resistant bacterial species that are a major threat to global health and require urgent development of novel therapeutics against them: *Enterococcus faecium*, *Staphylococcus aureus*, *Klebsiella pneumoniae*, *Acinetobacter baumannii*, *Pseudomonas aeruginosa*, and *Enterobacter* (Lakemeyer *et al.*, 2018).

To handle the issue of antibiotic resistance, the development of novel antibiotics is necessary (Aslam *et al.*, 2018; Blair, 2018; Ventola, 2015; Lakemeyer *et al.*, 2018; Spaulding *et al.*, 2018). In comparison to the rate of development of antimicrobial resistance, the pace of discovery or development of new antibiotics is very slow: in the last 2 decades only two new classes of antibiotics were introduced for clinical use (Lakemeyer *et al.*, 2018). Consequently, the use of vaccines, lysins, antibodies, probiotics, bacteriophages and antimicrobial peptides (AMPs) is becoming popular in therapeutics as alternatives to antibiotics (Aslam *et al.*, 2018). For designing new drugs, the use of AMPs is rapidly gaining attention (Aslam *et al.*, 2018; Kampshoff *et al.*, 2019; Costa *et al.*, 2019; Yu *et al.*, 2018). AMPs exhibit different biological activities against microbes, e.g., bacteria, viruses, fungi, etc. (Aslam *et al.*, 2018), have higher inhibition rates than antibiotics, and can potentially slow down the evolution of antibiotic resistance as well (Yu *et al.*, 2018).

Potential AMP candidates need to be tested and evaluated experimentally before entering clinical trials. The prediction of AMPs using machine learning techniques reduces the cost of identifying the effectiveness of a peptide sequence against microbial species in the wet lab by pre-screening potential antimicrobial peptides. A number of machine learning based AMP predictors are available in the literature (Gull *et al.*, 2019; Bhadra *et al.*, 2018; Torrent *et al.*, 2009; Waghu *et al.*, 2015; Lin and Xu, 2016; Agrawal and Raghava, 2018). The primary issue with these *un-targeted* predictors is that they are unable to predict whether a given peptide sequence will be effective against a given target microbial species or not (see Fig 1). Only a small number of targeted predictors exist in the literature but they are not able to generate predictions for novel microbial species (Kleandrova *et al.*, 2016; Vishnepolsky *et al.*, 2018; Speck-Planche *et al.*, 2016). Vishnepolsky et al. developed a predictor for 6 different gram-negative bacterial strains (Vishnepolsky *et al.*, 2018). The AMP predictor by Kleandrova et al. used 70 different gram-negative strains of bacteria in training to predict antimicrobial and cytotoxic activity of individual amino acids in a peptide sequence for different strains (Kleandrova *et al.*, 2016). Although they covered a large set of bacterial species, their method can generate predictions for only specific strains of gram-negative bacterial strains. Unavailability of their predictor for public use is also a limitation (Kleandrova *et al.*, 2016). The major drawback in targeted predictors is their inability of predicting a peptide's antimicrobial activity for novel microbial species. The prediction of antimicrobial activity of a peptide without knowing the microbial species against which the peptide is effective is not meaningful.

**Table 1. Filtering criteria applied to DBAASP database to obtain required dataset**

| Filtering criteria | Number of peptides |
|---|---|
| DBAASP monomer peptides | 12,984 |
| Sequences with length >5 | 12,517 |
| Sequences with microbial targets (excluding cancers) | 9,890 |
| Sequences with MIC in ($\mu M$) or ($\mu g/mL$) | 8,045 |
| Sequences with target species genomes available in NCBI (Coordinators, 2016) | 8,025 |
| Sequences with at least one target species with MIC $\leq$ 25 $\mu g/mL$ | 5,710 |

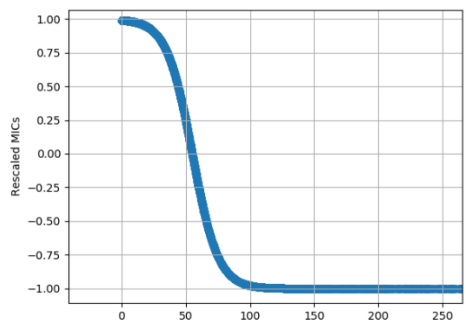

**Fig 2.** MICs converted to continuous labels between -1 to +1 using bipolar sigmoid function

In this work, we have developed a machine learning model to overcome this limitation. The proposed model takes amino acid sequence of a peptide and the genomic sequence of a target microbial species to predict the effectiveness of the peptide against that species in a targeted manner.

## 2 Methods

### 2.1. Data collection and preprocessing

For constructing the dataset used for training and evaluation of our machine learning models, we have used DBAASP version 2 (Pirtskhalava *et al.*, 2015). DBAASP has been widely used in recent studies in this field (Kleandrova *et al.*, 2016; Youmans *et al.*, 2017; Vishnepolsky *et al.*, 2018; Speck-Planche *et al.*, 2016; Win *et al.*, 2017). It contains a total of 12, 984 peptide sequences and their experimentally verified minimum inhibitory concentrations (MICs) against various target microbial species. In order to construct our dataset from DBAASP, we have used peptides with length greater than 5 amino acids whose experimentally validated MICs are available in micro molar ($\mu M$) or microgram per milliliter ($\mu g/mL$). We also ensured that the genomes of the target species are available in NCBI (Coordinators, 2016) and that each peptide in our dataset has at least one target species for which its MIC was $\leq$ 25 $\mu g/mL$ (Vishnepolsky *et al.*, 2018). The details of different filtration stages to extract the dataset of our interest are given in Table-1. DBAASP reports the effectiveness of a peptide sequence against multiple strains of a microbial species. We have taken the minimum MIC of a peptide across different strains of a species as its MIC against that species. All MIC values have been converted to $\mu g/mL$ (Kleandrova *et al.*, 2016). Our final dataset comprises of 5,710 peptides that are effective against a total of 336 different microbial species. The details of individu-al peptides and their MICs against their target species is given in supplementary material.

As an additional preprocessing step, we have scaled the MIC scores using a sigmoidal curve such that MIC scores $\leq 25\ \mu g/mL$ are mapped onto +1 and those $\geq 100\ \mu g/mL$ are mapped to -1 (see Fig. 2). For this purpose, we have utilized a sigmoid rescaling function which maps raw MIC scores $y$ as follows:

$$y' = s\left(-\frac{y-55}{10}\right) \text{ with } s(z) = 2\left(\frac{e^z}{1+e^z}\right) - 1.$$

This rescaling ensures that subsequent processing and machine learning models are not affected by large variations in MICs across different target species and peptides which can vary from a few $\mu g/mL$ to more than 2000 $\mu g/mL$. If the MIC of a peptide is not known for a species, its rescaled score is set at 0.0.

## 2.2. Feature extraction

To predict antimicrobial activity of a peptide against given species through machine learning, we need features of peptide and genomic sequence of target microbial species as discussed below (see Fig. 3).

### 2.2.1 Amino Acid Sequence features

In order to obtain peptide-level features, we have used one-hot encoding of the peptide sequence that results in a 40-dimensional feature vector (frequency count of 20 L-amino acids and 20 D-amino acids). The feature representation models the type of amino acid (L and D) in the peptide sequence separately as peptide bioactivity is dependent upon the type of amino acids (Cava *et al.*, 2011; Mangoni *et al.*, 2006; Baltz, 2009; Kawai *et al.*, 2004). The resulting feature vectors for a given peptide is normalized to unit norm. We have also analyzed 2-mer composition which results in a $40^2 = 1600$-dimensional feature vector (Leslie *et al.*, 2001).

DBAASP (Pirtskhalava *et al.*, 2015) also provides information about N-terminus and C-terminus modifications of peptides which can play a significant role in their antimicrobial activity. Modification at N-terminus and C-terminus of peptides can change their biological activity (Crusca Jr *et al.*, 2011). We have used one-hot encoding to capture information about C- and N-terminus modifications in our feature representation. The sequence features are concatenated with C and N termini features. Details about the different types of C and N termini modifications are given in supplementary information.

### 2.2.2 Genomic features

In order to perform targeted prediction of antimicrobial activity of a peptide sequence against a particular species through machine learning, we need to extract species-level features as well. The literature reports the use of mono, di, tri and tetra-nucleotide composition of genomic sequences for comparison or clustering of genomes (Karlin and Ladunga, 1994; Karlin *et al.*, 1998; Kariin and Burge, 1995; Karlin, 1998; Nakashima *et al.*, 1997, 1998; Pride *et al.*, 2003; Takahashi *et al.*, 2009). As a consequence, we have extracted features from complete genomes of species downloaded from NCBI (Coordinators, 2016). For feature extraction the counts of 1-mer, 2-mer, 3-mer and 4-mer are calculated from a given genome sequence and normalized to unit norm resulting in a 340-dimensional feature representation.

## 2.3. Prediction Models

To predict whether a given peptide sequence will be effective against a target microbial species or not, we have proposed a zero-shot machine learning model. We compare the proposed model to a conventional ma chine learning model as a baseline as discussed below. In order to aid the reader in understanding our modeling approach for baseline and zero-shot predictors, we denote a peptide sequence by its $d$-dimensional feature vector $x_i$, $i = 1, \ldots, 5710$ whereas a particular microbial species is represented by an $a$-dimensional attribute vector $s_j$, for $j = 1, \ldots, 336$ based on its genomic sequence. We denote the rescaled MIC of a peptide $x_i$ against species $s_j$ by the target variable $y_{ij}$. The prediction problem can then be expressed as finding a mathematical function $f(x_i, s_j; \Theta)$ parameterized by learnable parameters $\Theta$ that can predict the effectiveness of a sequence $x_i$ for microbial species $s_j$.

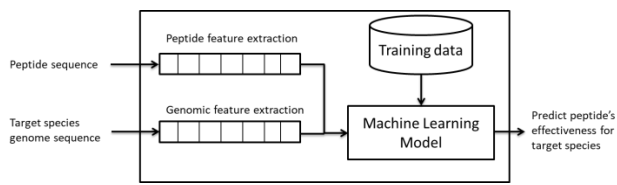

**Fig 3.** Proposed model framework using features of peptide and genomic sequences

### 2.3.1 Baseline models

We have chosen Radial Basis Function SVM (Cortes and Vapnik, 1995) and XGBoost (Chen and Guestrin, 2016) as baseline models due to their widespread use and ease of modeling. For this purpose, in order to predict the effectiveness of a given peptide sequence against a microbial species, we construct a joint feature representation $\phi_{ij} = \begin{bmatrix} x_i \\ s_j \end{bmatrix}$ by concatenating peptide and species level features with the associated training label $y_{ij}$ set to +1 (antimicrobial) if the MIC of peptide $x_i$ for species $s_j$ is $\leq 25\ \mu g/mL$ and -1 (non-antimicrobial) if the MIC is $\geq 100\ \mu g/mL$. A conventional SVM or XGBoost model can then be trained over such a data set.

### 2.3.2 Zero and Few shot learning

In this work, we propose to model the problem of targeted antimicrobial activity prediction through zero shot learning (ZSL) (Romera-Paredes and Torr, 2015). Widely used in object classification and computer vision, ZSL allows a classification model to generate predictions for novel classes which were not available at training time (Socher *et al.*, 2013; Norouzi *et al.*, 2013; Fu *et al.*, 2015). This is achieved by learning the definition of a class through an attribute vector representation instead of predicting class labels directly as in conventional classification. Many variants of ZSL have been proposed in the literature (Palatucci *et al.*, 2009; Zhang and Saligrama, 2015; Socher *et al.*, 2013; Norouzi *et al.*, 2013; Fu *et al.*, 2015; Kodirov *et al.*, 2017; Romera-Paredes and Torr, 2015). While ZSL assumes that no examples of a novel class presented during testing are available for training, the related case of few-shot learning aims at building a machine learning model such that only a few training examples are available for the target class (Snell *et al.*, 2017; Sung *et al.*, 2018; Gidaris and Komodakis, 2018; Garcia and Bruna, 2017; Ravi and Larochelle, 2016). Few Shot Learning (FSL) techniques perform significantly better than conventional classification methods when the number of training examples is very small (Snell *et al.*, 2017; Sung *et al.*, 2018; Gidaris and Komodakis, 2018).

The problem of targeted antimicrobial activity prediction is ideally suited to zero and few shot learning: in typical machine learning guided design of wet lab experiments for screening potential peptides that are effective against a target microbial species, no or very few peptides with known labels are available for training. Furthermore, in order to predict how effective a peptide is against a novel microbial species for which no or very few training examples are available, we can model the target microbial species as a class represented by an attribute vector based on its genomic sequence. In this work, we have used the ZSL scheme given by Romera-Paredes and Torr (Romera-Paredes and Torr, 2015). For predicting the MIC of a peptide sequence for a target species, the discriminant function used by the ZSL model of Romera-Paredes and Torr (Romera-Paredes and Torr, 2015) can be written as $f(x_i, s_j; \Theta) = x_i^T \Theta s_j$ with the learnable weight matrix $\Theta \in \mathbb{R}^{d \times a}$. If the number of peptides and species (classes) available during training are $m$ and $z$, respectively and the rescaled MIC scores for each of the peptide against each microbe is represented by the $m \times z$ matrix $Y \in [-1,1]^{m \times z}$, the learning problem for ZSL can be formulated as the following optimization problem:

$$\Theta^* = \underset{\Theta \in \mathbb{R}^{d \times a}}{\mathrm{argmin}} \; \|X^T \Theta S - Y\|_{Fro}^2 + (\gamma \|\Theta S\|_F^2 + \lambda \|X^T \Theta\|_F^2 + \gamma \lambda \|\Theta\|_F^2)$$

Here, $X \in \mathbb{R}^{d \times m}$ and $S \in \mathbb{R}^{a \times z}$ represent matrices of all peptide features ($m$ examples each with a $d$-dimensional feature vector) and attributes of microbial species ($z$ classes each with $a$ attributes), respectively. The first term represents the loss function with the aim of minimizing the error between predicted and target MICs. The second term $(\gamma \|\Theta S\|_F^2 + \lambda \|X^T \Theta\|_F^2 + \gamma \lambda \|\Theta\|_F^2)$ is the regularization factor that ensures smoothness of the prediction function $f(x, s; \Theta)$ and sparsity of the weight matrix $\Theta$ through penalization of the Frobenius norm $\|\cdot\|_F^2$ of respective matrices. $\gamma$ and $\lambda$ are regularization hyper-parameters. In addition to better performance over benchmark datasets, another reason for choosing this ZSL implementation is the existence of a computationally efficient closed-form solution of its underlying optimization problem which can be written as follows:

$$\Theta^* = (XX^T + \gamma I)^{-1} XYS^T (SS^T + \lambda I)^{-1}$$

Once the optimal weight matrix $\Theta^*$ has been obtained, the predictions for a peptide (represented by the feature vector $x$) for species (represented by the attribute vector $s$) can be generated by the decision function $f(x, s; \Theta^*) = x^T \Theta^* s$. Note that this decision function can be used for generating predictions both for novel peptides and novel species provided their attribute representation $s$ is available. The most likely target species for a given peptide can be identified by simply ranking the resulting decision function scores across a given list of potential target species.

This formulation can be kernelized for non-linear kernels as well by applying the Representer theorem to the underlying optimization problem (Romera-Paredes and Torr, 2015). For this purpose, an $m \times m$ sized kernel matrix $K$ with $K_{ij} = k(x_i, x_j)$ is computed over the training data using a kernel function such as the radial basis function (RBF) $k(a, b) = exp(-\kappa \|a - b\|^2)$ with the hyperparameter $\kappa > 0$. The closed form solution of the kernelized ZSL optimization problem requires calculation of an $m \times a$ sized instance-attribute association matrix $\mathbf{A}$ from training data as follows (see (Romera-Paredes and Torr, 2015) for details):

$$A = (K^T K + \gamma I)^{-1} KYS (S^T S + \lambda I)^{-1}$$

For inference or prediction of effectiveness of a peptide represented by a feature vector $x$ against a microbial species represented by its attribute vector $s$, an $m$-dimensional vector of kernel scores $k(x) = [k(x, x_1) \; k(x, x_2) \; \cdots \; k(x, x_m)]^T$ of the test example with each training example is computed and used in the kernelized prediction function $f(x, s; A) = k(x)^T A s$.

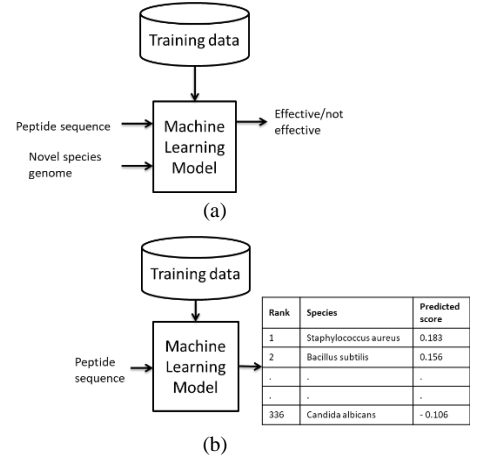

**Fig 4.** (a) PAP takes inputs of a peptide sequence and a novel species genome to predict whether a peptide is effective against a given species or not; (b) TSR requires a novel peptide sequence and predicts the microbe that is most likely to be targeted by that peptide (out of 336 given species)

It is important to note that this framework extends seamlessly to FSL by simply adding further training instances for a target class. The hyperparameters of the model $(\gamma, \lambda, \kappa)$ are tuned through cross-validation. The best performance of the model was found using $\gamma = 2.0$, $\lambda = 0.0001$, and the hyperparameter $\kappa$ of RBF kernel is set to 2.0.

### 2.4. Performance evaluation

We consider two practical use-cases of our system: 1) Target Species Ranking (TSR): given a set of microbial species for which labeled peptide sequences are available for training, predict the microbe that is most-likely to be targeted by a novel peptide sequence and, 2) Peptide Activity Prediction for Novel Species (PAP): predict whether a peptide is effective against a given species or not such that no or very few peptide examples for that species are available during training (i.e., Zero Shot or Few Shot Learning) (see Fig. 4). It is important to note that both these scenarios reflect practical use cases for biologists who are interested in machine-learning guided discovery for targeted antimicrobial peptides.

In order to evaluate the performance of baseline and proposed machine learning models for TSR, we have used 5-fold cross validation (John Lu, 2010). The dataset of 5,710 peptides is divided into 5 non-overlapping folds. A given model is trained on labeled examples of all peptides in 4 folds and tested on the remaining peptides. This process is repeated 5 times, once for each fold. For each test peptide in a fold, model scores for all 336 species are sorted in descending order. The rank of the highest scoring microbe that is a known target of the given test peptide (positive example) is used as a peptide-specific performance metric. This simple biologist-centric performance metric called Rank of First Positive Prediction (RFPP) is based on the premise that an ideal machine learning model should assign high score to a known target species of a given peptide sequence and, consequently, rank target species at lower ranks in the sorted list in comparison to non-target species (Minhas *et al.*, 2014). As a result, for an ideal machine learning model, the RFPP for all test peptides should be 1.0. As discussed in the results section, we report the percentile-wise RFPP scores for all test peptides for different machine learning models together with a random predictor as experimental control. The RFPP score at a certain percentile $p$, henceforth denoted by $RFPP(p)$ is defined as follows: $RFPP(p) = q$, if $p\%$ test peptides have at least one known target microbial species among their top $q$ predictions (out of 336). Thus, for an ideal classifier $RFPP(100) = 1$, i.e., for every peptide, the top scoring species is a real target species of the given test peptide. RFPP is a biologist-centric metric as it tells us directly how often top-ranking predictions of a peptide can be expected to correspond to true target species and it can be directly used in experiment design.

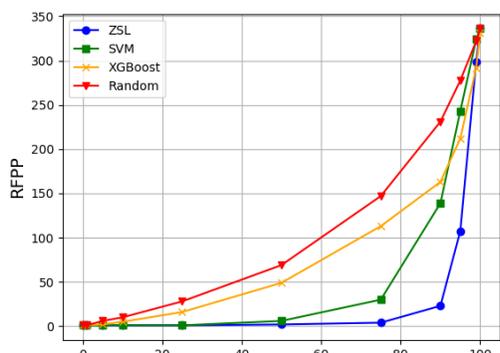

**Fig 5.** Percentile-wise RFPP for target species ranking for various machine learning models

For PAP, i.e., predicting a peptide's effectiveness for a novel species, our proposed modeling approach takes peptide and genomic sequences as input and the score generated by the decision function of a machine learning model is used for classification of peptide sequences for individual species. In order to quantify predictive accuracy, a selected set of 17 test species from DBAASP with a small but sufficient number (75-180) of known positive and negative peptide examples is used (details given in Table-2). For ZSL, the model is trained on all examples from other species and its predictive performance is evaluated for individual species in Table 2 using area under the receiver operating characteristic curve (AUC-ROC) as a performance metric (Davis and Goadrich, 2006). For few shots learning (FSL), a few positive and negative examples of a test species (1, 2, 4, 8 and half of all available examples for that species) are randomly sampled for training together with all examples from all other species and the model is evaluated on the remaining examples of the test species. This process is repeated 20 times with different species-level training and test examples to get average AUC-ROC scores and their standard deviation.

## 3    Results

In this section, we discuss the results for the two learning tasks below.

### 3.1    Target Species Ranking (TSR)

Fig 5 shows the percentile-wise RFPP scores for all classifiers. As discussed in section 2.4, the ideal RFPP score for all peptides is 1.0. For the random classifier that generates a random score for a given example, the median RFPP is 75, i.e., for 50% test peptides in cross-validation, a true target species is within the top 75 (out of 336) predictions. In contrast, for XGBoost and SVM baseline models, the median RFPPs are 50 and 10, respectively. However, the proposed model performs much better than these baseline models: the RFPP for the proposed model at the 75[th] percentile is 1.0, i.e., the for up to 75% peptides, the top prediction by the model is correct. This clearly shows the effectiveness of the proposed prediction scheme for identifying the correct target species of a peptide. For TSR use case optimal results of SVM and XGBoost were obtained with 2-mer composition features of peptides.

### 3.2    Peptide Activity Prediction for Novel Species

Table 2 shows the results of various machine learning models for the Peptide Activity Prediction (PAP) task. In this task the objective is to evaluate whether a given machine learning model can correctly predict peptides that target a novel species for which none or very few training examples are available. For this purpose, we compare the performance of conventional machine learning models (SVM, XGBoost), the proposed Zero Shot Learning (ZSL) and Few Shot Learning (FSL) models in addition to existing state of the art non-targeted antimicrobial activity predictors (CAMP (Waghu et al., 2015) (Gabere and Noble, 2017) and AMAP (Gull et al., 2019)). For this use case, XGBoost with amino acid composition features performed significantly better than SVM (results not shown for brevity). However, the prediction performance of XGBoost was typically no better than a random classifier especially when the number of training examples from a given test species was very small (see Supplementary Information for complete results). Similarly, existing state of the art methods such as CAMP (Waghu et al., 2015) and AMAP (Gull et al., 2019) do not give satisfactory predictive performance for the chosen species. In contrast, the proposed few shot learning model performs significantly better than other methods with an expected increase in prediction accuracy when the number of training examples of a species is increased.

### 3.3    Webserver and Code

The webserver developed for proposed model together with the code is available at the URL:http://ampzero.pythonanywhere.com. The webserver takes a peptide sequences in FASTA format along with any C-terminus and N-terminus modifications as input together with the genome of a species in order to predict the degree of effectiveness of the peptide against the given species. Additionally, the user can upload a list of known positive and negative example peptide sequences for the given species for generating few shot learning based predictions.

## 4    Conclusions

We have developed a targeted antimicrobial activity predictor called AMPZero can predict the effectiveness of a given peptide sequence against a given target species. The use of zero and few shot learning in the proposed model helps in overcoming the shortcomings of conventional machine learning techniques for this purpose. Our cross-validation analysis shows that the proposed model can perform better than existing approaches and it can be easily integrated in experimental discovery of antimicrobial peptide sequences for novel species.

## Acknowledgements

Sadaf Gull is supported by a grant under indigenous 5000 Ph.D. fellowship scheme by the Higher Education Commission (HEC) of Pakistan.

*Conflict of Interest:* none declared.

**Table 2** Results for Peptide Activity Prediction for Novel Species. The first column indicates the type of the different test species used in this analysis. The species name together with the total number of positive (P) and negative (N) examples available for that species are given in the second column. Results for zero shot learning (ZSL) in which no examples of the given test species are included in training are shown for the proposed ZSL model. For few shot learning results for different number of training examples (1, 2, 4, 8 and Half of all available examples) of the target species are shown. In the interest of relevance and brevity results for XGBoost are shown only when half of the available examples are used for training. CAMP and AMAP are existing state of the art predictors for antimicrobial activity and the prediction results were obtained using their respective webservers. Values in bold indicate the highest prediction performance. Note that the average AUC-ROC across multiple runs is reported together with the standard deviation (in parenthesis).

| Species Type | Species Name ↓ | Machine Learning Models | | | | | | | | |
|---|---|---|---|---|---|---|---|---|---|---|
| | | ZSL | FSL | | | | | XGBoost | CAMP | AMAP |
| | No. of Tr. Examples → | 0 | 1 | 2 | 4 | 8 | Half | Half | | |
| *Fungus* | *Aspergillus fumigatus* | 0.746 | 0.807 | 0.806 | 0.820 | 0.835 | **0.882** | 0.614 | 0.798 | 0.545 |
| | (P: 44, N: 33) | (0.056) | (0.059) | (0.041) | (0.046) | (0.043) | (0.043) | (0.073) | (0.051) | (0.055) |
| | *Candida glabrata* | 0.652 | 0.594 | 0.620 | 0.628 | 0.691 | **0.781** | 0.677 | 0.350 | 0.489 |
| | (P35: , N:47 ) | (0.056) | (0.087) | (0.084) | (0.088) | (0.072) | (0.052) | (0.087) | (0.047) | (0.081) |
| | *Candida parapsilosis* | 0.430 | 0.473 | 0.507 | 0.562 | 0.663 | **0.789** | 0.639 | 0.660 | 0.662 |
| | (P:51 , N:33 ) | (0.088) | (0.094) | (0.106) | (0.093) | (0.089) | (0.055) | (0.120) | (0.069) | (0.075) |
| | *Candida tropicalis* | 0.755 | 0.712 | 0.735 | 0.771 | 0.803 | **0.865** | 0.669 | 0.703 | 0.561 |
| | (P:88 , N:16 ) | (0.078) | (0.072) | (0.076) | (0.078) | (0.071) | (0.042) | (0.066) | (0.076) | (0.058) |
| | *Cryptococcus neoformans* | 0.504 | 0.497 | 0.487 | 0.518 | 0.628 | **0.628** | 0.541 | 0.576 | 0.581 |
| | (P:167 , N:14 ) | (0.102) | (0.110) | (0.104) | (0.103) | (0.068) | (0.068) | (0.078) | (0.089) | (0.084) |
| | *Saccharomyces cerevisiae* | 0.405 | 0.627 | 0.634 | 0.650 | 0.681 | **0.788** | 0.604 | 0.388 | 0.448 |
| | (P:132 , N:36 ) | (0.061) | (0.064) | (0.060) | (0.064) | (0.072) | (0.052) | (0.046) | (0.053) | (0.043) |
| | *Fusarium oxysporum* | 0.856 | 0.914 | 0.919 | 0.932 | 0.943 | **0.961** | 0.696 | 0.418 | 0.396 |
| | (P:125 , N:18 ) | (0.045) | (0.040) | (0.043) | (0.038) | (0.033) | (0.021) | (0.094) | (0.049) | (0.033) |
| Gram Negative Bacteria | *Enterobacter aerogenes* | 0.468 | 0.588 | 0.599 | 0.646 | 0.731 | **0.826** | 0.773 | 0.550 | 0.443 |
| | (P:36 , N:49 ) | (0.061) | (0.084) | (0.063) | (0.088) | (0.078) | (0.051) | (0.069) | (0.067) | (0.069) |
| | *Erwinia amylovora* | 0.478 | 0.385 | 0.450 | 0.543 | 0.714 | 0.892 | **0.907** | 0.877 | 0.385 |
| | (P112: , N:35 ) | (0.059) | (0.062) | (0.068) | (0.069) | (0.069) | (0.047) | (0.032) | (0.021) | (0.046) |
| | *Pasteurella multocida* | 0.722 | 0.745 | 0.807 | 0.876 | 0.924 | **0.957** | 0.914 | 0.528 | 0.295 |
| | (P:37 , N:53 ) | (0.052) | (0.088) | (0.080) | (0.042) | (0.026) | (0.019) | (0.046) | (0.067) | (0.039) |
| | *Proteus mirabilis* | 0.714 | 0.729 | 0.733 | 0.748 | 0.767 | **0.836** | 0.731 | 0.377 | 0.269 |
| | (P:27 , N:105 ) | (0.052) | (0.045) | (0.047) | (0.054) | (0.045) | (0.048) | (0.079) | (0.046) | (0.065) |
| | *Proteus vulgaris* | 0.710 | 0.780 | 0.794 | 0.818 | 0.840 | **0.909** | 0.667 | 0.465 | 0.567 |
| | (P:84 , N:34 ) | (0.038) | (0.050) | (0.050) | (0.050) | (0.045) | (0.030) | (0.071) | (0.048) | (0.048) |
| | *Serratia marcescens* | 0.782 | 0.843 | 0.864 | 0.883 | 0.886 | **0.921** | 0.571 | 0.397 | 0.418 |
| | (P:48 , N:62 ) | (0.040) | (0.042) | (0.031) | (0.028) | (0.039) | (0.021) | (0.068) | (0.084) | (0.046) |
| Gram Positive Bacteria | *Listeria innocua* | 0.686 | 0.688 | 0.710 | 0.738 | 0.763 | **0.804** | 0.672 | 0.371 | 0.402 |
| | (P:64 , N:36 ) | (0.038) | (0.067) | (0.064) | (0.056) | (0.062) | (0.057) | (0.052) | (0.048) | (0.032) |
| | *Streptococcus mutans* | 0.437 | 0.570 | 0.597 | 0.616 | 0.825 | **0.825** | 0.767 | 0.472 | 0.706 |
| | (P:129 , N:11 ) | (0.119) | (0.118) | (0.116) | (0.136) | (0.045) | (0.045) | (0.108) | (0.099) | (0.083) |
| | *Streptococcus pneumoniae* | 0.591 | 0.619 | 0.609 | 0.622 | 0.623 | **0.701** | 0.507 | 0.3081 | 0.384 |
| | (P:86 , N:17 ) | (0.077) | (0.090) | (0.089) | (0.079) | (0.077) | (0.066) | (0.071) | (0.057) | (0.077) |
| | *Streptococcus pyogenes* | 0.660 | 0.733 | 0.737 | 0.747 | 0.873 | **0.873** | 0.669 | 0.569 | 0.737 |
| | (P:161 , N:09 ) | (0.050) | (0.044) | (0.044) | (0.053) | (0.037) | (0.037) | (0.069) | (0.156) | (0.062) |

**References**

Afsar Minhas,F. ul A. *et al.* (2014) PAIRpred: Partner-specific prediction of interacting residues from sequence and structure. *Proteins: Structure, Function, and Bioinformatics*, **82**, 1142–1155.

Agrawal,P. and Raghava,G.P. (2018) Prediction of Antimicrobial Potential of a Chemically Modified Peptide From Its Tertiary Structure. *Frontiers in Microbiology*, **9**, 2551.

Aslam,B. *et al.* (2018) Antibiotic resistance: a rundown of a global crisis. *Infection and drug resistance*, **11**, 1645.

Baltz,R.H. (2009) Daptomycin: mechanisms of action and resistance, and biosynthetic engineering. *Current opinion in chemical biology*, **13**, 144–151.

Bhadra,P. *et al.* (2018) AmPEP: Sequence-based prediction of antimicrobial peptides using distribution patterns of amino acid properties and random forest. *Scientific reports*, **8**, 1697.

Blair,J.M. (2018) A climate for antibiotic resistance. *Nature Climate Change*, **8**, 460.

Cava,F. *et al.* (2011) Emerging knowledge of regulatory roles of D-amino acids in bacteria. *Cellular and Molecular Life Sciences*, **68**, 817–831.

Chen,T. and Guestrin,C. (2016) Xgboost: A scalable tree boosting system. In, *Proceedings of the 22nd acm sigkdd international conference on knowledge discovery and data mining*. ACM, pp. 785–794.

Coordinators,N.R. (2016) Database resources of the national center for biotechnology information. *Nucleic acids research*, **44**, D7.

Cortes,C. and Vapnik,V. (1995) Support-vector networks. *Machine learning*, **20**, 273–297.

Costa,F. *et al.* (2019) Clinical Application of AMPs. In, *Antimicrobial Peptides*. Springer, pp. 281–298.

Crusca Jr,E. *et al.* (2011) Influence of N-terminus modifications on the biological activity, membrane interaction, and secondary structure of the antimicrobial peptide hylin-a1. *Peptide Science*, **96**, 41–48.

Davis,J. and Goadrich,M. (2006) The relationship between Precision-Recall and ROC curves. In, *Proceedings of the 23rd international conference on Machine learning*. ACM, pp. 233–240.

Fu,Y. *et al.* (2015) Transductive multi-view zero-shot learning. *IEEE transactions on pattern analysis and machine intelligence*, **37**, 2332–2345.

Gabere,M.N. and Noble,W.S. (2017) Empirical comparison of web-based antimicrobial peptide prediction tools. *Bioinformatics*, **33**, 1921–1929.

Garcia,V. and Bruna,J. (2017) Few-shot learning with graph neural networks. *arXiv preprint arXiv:1711.04043*.

Gidaris,S. and Komodakis,N. (2018) Dynamic few-shot visual learning without forgetting. In, *Proceedings of the IEEE Conference on Computer Vision and Pattern Recognition*., pp. 4367–4375.

Gull,S. *et al.* (2019) AMAP: Hierarchical multi-label prediction of biologically active and antimicrobial peptides. *Computers in biology and medicine*, **107**, 172–181.

John Lu,Z. (2010) The elements of statistical learning: data mining, inference, and prediction. *Journal of the Royal Statistical Society: Series A (Statistics in Society)*, **173**, 693–694.

Kampshoff,F. *et al.* (2019) A Pilot Study of the Synergy between Two Antimicrobial Peptides and Two Common Antibiotics. *Antibiotics*, **8**, 60.

Kariin,S. and Burge,C. (1995) Dinucleotide relative abundance extremes: a genomic signature. *Trends in genetics*, **11**, 283–290.

Karlin,S. *et al.* (1998) Comparative DNA analysis across diverse genomes. *Annual review of genetics*, **32**, 185–225.

Karlin,S. (1998) Global dinucleotide signatures and analysis of genomic heterogeneity. *Current opinion in microbiology*, **1**, 598–610.

Karlin,S. and Ladunga,I. (1994) Comparisons of eukaryotic genomic sequences. *Proceedings of the National Academy of Sciences*, **91**, 12832–12836.

Kawai,Y. *et al.* (2004) Structural and functional differences in two cyclic bacteriocins with the same sequences produced by lactobacilli. *Appl. Environ. Microbiol.*, **70**, 2906–2911.

Kleandrova,V.V. *et al.* (2016) Enabling the discovery and virtual screening of potent and safe antimicrobial peptides. simultaneous prediction of antibacterial activity and cytotoxicity. *ACS combinatorial science*, **18**, 490–498.

Kodirov,E. *et al.* (2017) Semantic autoencoder for zero-shot learning. In, *Proceedings of the IEEE Conference on Computer Vision and Pattern Recognition*., pp. 3174–3183.

Lakemeyer,M. *et al.* (2018) Thinking Outside the Box—Novel Antibacterials To Tackle the Resistance Crisis. *Angewandte Chemie International Edition*, **57**, 14440–14475.

Leslie,C. *et al.* (2001) The spectrum kernel: A string kernel for SVM protein classification. In, *Biocomputing 2002*. World Scientific, pp. 564–575.

Lin,W. and Xu,D. (2016) Imbalanced multi-label learning for identifying antimicrobial peptides and their functional types. *Bioinformatics*, **32**, 3745–3752.

Mangoni,M.L. *et al.* (2006) Effect of natural L-to D-amino acid conversion on the organization, membrane binding, and biological function of the antimicrobial peptides bombinins H. *Biochemistry*, **45**, 4266–4276.

Nakashima,H. *et al.* (1997) Di. erences in Dinucleotide Frequencies of Human, Yeast, and Escherichia coli Genes. *DNA Research*, **4**, 185–192.

Nakashima,H. *et al.* (1998) Genes from nine genomes are separated into their organisms in the dinucleotide composition space. *DNA Research*, **5**, 251–259.

Norouzi,M. *et al.* (2013) Zero-shot learning by convex combination of semantic embeddings. *arXiv preprint arXiv:1312.5650*.

Palatucci,M. *et al.* (2009) Zero-shot learning with semantic output codes. In, *Advances in neural information processing systems*., pp. 1410–1418.

Pirtskhalava,M. *et al.* (2015) DBAASP v. 2: an enhanced database of structure and antimicrobial/cytotoxic activity of natural and synthetic peptides. *Nucleic acids research*, **44**, D1104–D1112.

Pride,D.T. *et al.* (2003) Evolutionary implications of microbial genome tetranucleotide frequency biases. *Genome research*, **13**, 145–158.

Ravi,S. and Larochelle,H. (2016) Optimization as a model for few-shot learning.

Romera-Paredes,B. and Torr,P. (2015) An embarrassingly simple approach to zero-shot learning. In, *International Conference on Machine Learning*., pp. 2152–2161.

Snell,J. *et al.* (2017) Prototypical networks for few-shot learning. In, *Advances in Neural Information Processing Systems*., pp. 4077–4087.

Socher,R. *et al.* (2013) Zero-shot learning through cross-modal transfer. In, *Advances in neural information processing systems*., pp. 935–943.

Spaulding,C.N. *et al.* (2018) Precision antimicrobial therapeutics: the path of least resistance? *NPJ biofilms and microbiomes*, **4**, 4.

Speck-Planche,A. *et al.* (2016) First multitarget chemo-Bioinformatic model to enable the discovery of antibacterial peptides against multiple gram-positive pathogens. *Journal of chemical information and modeling*, **56**, 588–598.

Sung,F. *et al.* (2018) Learning to compare: Relation network for few-shot learning. In, *Proceedings of the IEEE Conference on Computer Vision and Pattern Recognition*., pp. 1199–1208.

Takahashi,M. *et al.* (2009) Estimation of bacterial species phylogeny through oligonucleotide frequency distances. *Genomics*, **93**, 525–533.

Torrent,M. *et al.* (2009) A theoretical approach to spot active regions in antimicrobial proteins. *BMC bioinformatics*, **10**, 373.

Ventola,C.L. (2015) The antibiotic resistance crisis: part 1: causes and threats. *Pharmacy and therapeutics*, **40**, 277.

Vishnepolsky,B. *et al.* (2018) Predictive Model of Linear Antimicrobial Peptides Active against Gram-Negative Bacteria. *Journal of chemical information and modeling*, **58**, 1141–1151.

Waghu,F.H. *et al.* (2015) CAMPR3: a database on sequences, structures and signatures of antimicrobial peptides. *Nucleic acids research*, **44**, D1094–D1097.

Win,T.S. *et al.* (2017) HemoPred: a web server for predicting the hemolytic activity of peptides. *Future medicinal chemistry*, **9**, 275–291.

Youmans,M. *et al.* (2017) Long short-term memory recurrent neural networks for antibacterial peptide identification. In, *2017 IEEE International Conference on Bioinformatics and Biomedicine (BIBM)*. IEEE, pp. 498–502.

Yu,G. *et al.* (2018) Predicting drug resistance evolution: insights from antimicrobial peptides and antibiotics. *Proceedings of the Royal Society B: Biological Sciences*, **285**, 20172687.

Zhang,Z. and Saligrama,V. (2015) Zero-shot learning via semantic similarity embedding. In, *Proceedings of the IEEE international conference on computer vision*., pp. 4166–4174.